%% file: fres.tex
\documentclass[usenatbib]{mn2e}
\usepackage{amsmath}
\usepackage{graphicx}
\usepackage{natbib}
\include{aas_journals}

\title[fractional excursion set theory]{
The fractional Brownian motion and the halo mass function}

\begin{document}
\author[J. Pan]{Jun Pan \thanks{jpan@pmo.ac.cn}\\
The Purple Mountain Observatory, 2 West Beijing Road, Nanjing 210008, China}

\maketitle

\begin{abstract}
The fractional Brownian motion with index $\alpha$ is introduced to construct 
the fractional excursion set model. A new mass function with single parameter $\alpha$
is derived within the formalism, of which the Press-Schechter 
mass function (PS) is a special case when $\alpha=1/2$. Although the new mass function
is computed assuming spherical collapse, comparison with the Sheth-Tormen fitting 
function (ST) shows that the new mass function of $\alpha\approx 0.435$ agrees with ST remarkably
well in high mass regime, while predicts more small mass halos than the ST but less than the PS. 
The index $\alpha$ is the Hurst exponent, which exact value in context of structure formation is
modulated by properties of the smoothing window function and the shape of power spectrum.
It is conjectured that halo merging rate and merging history in the fractional set theory might
be imprinted with the interplay between halos at small scales and their large scale 
environment. And the mass function in high mass regime can be a good tool to detect the 
non-Gaussianity of the initial density fluctuation.
\end{abstract}


%=========================================================================
\section{introduction}
Halo models are widely applied in the campaign of cosmological parameter estimation in precision
from cosmic large scale structures as well as in the expedition of understanding structure formation. 
The mass function is a fundamental ingredient of halo models. 
The most famous analytical formula of mass function, the Press-Schechter mass function (hereafter PS) was 
derived by \citet{PressSchechter1974} based on the spherical collapse model.
The PS function can be alternatively derived with the 
random walk or the excursion set formalism \citep{BondEtal1991}. By smoothing the linear density 
field on different scales with a sharp k-space filter, the density fluctuation within the characteristic 
scale could be regarded as a random walk against the variance of the smoothed field at this scale. 
Consequently the whole theory of random walk can be grafted to model the density contrast field.
The elegant theory provides a concise analytical framework to study various processes in cosmic 
structure formation, and is embraced with great interests by the community. For instance, 
\citet{LaceyCole1993} explicitly calculated the merger rate, halo 
formation time, and relevant properties of galaxy clusters; \citet{ShethTormen2002} adopted the excursion 
set theory with moving barrier to study ellipsoidal collapse of halos; \citet{ZhangHui2006} solved the 
excursion set theory with moving barrier of arbitrary shape and discussed the HII bubble size during 
re-ionization; and voids phenomenon is explored within the framework by \citet{FurlanettoPiran2006}.

The success of the random walk formalism in cosmology is prominent, but
the primary product of the excursion set theory, the PS mass function, is a poor description to simulations 
at all epochs \citep{ReedEtal2006}. The common practice is to parameterize the PS function, and 
then fit the function to simulations to pin down free parameters 
\citep[e.g.][]{ShethTormen1999, JenkinsEtal2001, ReedEtal2003, WarrenEtal2006}.
Many functional forms have been proposed by various authors to
account for different effects of ellipsoidal collapse \citep{ShethTormen2002}, 
angular momentum \citep{DelPopolo2006a} and the index of power spectrum \citep{ReedEtal2006}.
\citet{BetancortMontero2006} claims that the ``all-mass-at-center'' problem shall be
properly formulated to obtain the correct mass function in high mass regime, \citet{Lee2006}
assumes there is a break in the hierarchical merging process and obtains much shallower mass function
in low mass regime.

In this report, we construct a fractional excursion set theory by replacing the conventional
random walk with the fractional Brownian motion of index $\alpha$. The standard excursion set theory is 
simply a special case of the the new theory. The difference between the normal random walk and the fractional 
random walk lies in that the latter takes the correlation between walking steps into account. A new
 mass function is derived with the fractional excursion set theory, which contains one parameter 
$\alpha$ in connection with the correlation of steps of the random walk. 
Although the new mass function is derived with the
boundary condition of a single fixed absorbing barrier, i.e. in spherical collapse scenario, 
it is in good agreement with the Sheth-Tormen formula 
\citep[][hereafter ST]{ShethTormen1999} with $\alpha\approx0.435$ in high mass regime, while has more
small mass halos than ST and less small mass halos than PS. 

The layout of this paper is that at first we recite the excursion set theory briefly in Section 2, 
then in Section 3 we introduce the fractional Brownian motion to develop the fractional excursion set theory 
and subsequently derive a new mass function, and the last section is of discussion.

%===============================================================================
\section{the excursion set theory}
The initial density fluctuation $\delta=\rho/\bar\rho -1\ll 1$ in early universe is Gaussian
and evolves linearly. 
If the density contrast in a region exceeds a critical value $\delta_c$, the mass in that region will
collapse and be virialised in future to form a halo. 

As pointed out by \citet{BondEtal1991}, at an arbitrary point in the universe, the density contrast
smoothed with a window function $W_M(R)$ of characteristic scale $R$ is a function of the underlying 
total mass $M(R)\sim \bar\rho R^3$ included by the smoothing window, the $M$ effectively 
represents the scale R. The variation of the smoothed density contrast $\delta(M)$ forms a trajectory
in the plane of $\delta(M)$-$M$. The collapsing condition $\delta_c$ is turned into an absorbing barrier 
over the trajectory, at the largest $M$ where $\delta(M)$ firstly crosses 
the barrier, the trajectory will be absorbed, i.e. an object will form. 
The task to find how many objects will form in mass range ($M, M+dM$) is converted to the problem 
of tracing the fraction of trajectories passing through the barrier.

A quantity used to represent the smoothing scale in stead of the mass $M$ is the 
variance of the smoothed field
\begin{equation}
S(M)=\sigma^2(M)=\langle |\delta_M|^2 \rangle=\sum_k \langle |\delta_k|^2 \rangle\tilde{W}^2_M(k)\ ,
\end{equation}
where $\delta_k$ and $\tilde{W}_M(k)$ are the Fourier transform of $\delta$ and the window function $W_M(r)$
respectively. The smoothed density fluctuation can be written as 
\begin{equation}
\delta(S)=\delta(M)=\sum_{k} \delta_k \tilde{W}_M(k)\ ,
\end{equation}
which actually tells us that $\delta(S)$ is the sum of $\delta_k$ weighted by the window function
$\tilde{W}_M(k)$.
If the smoothing scale $R$ is sufficiently large, $S$ and $\delta(S)$ will be zero. Once we
decrease the smoothing scale $R$, since the window $W$ is a function of $R$, 
the weighting to Fourier modes of $\delta$ will change. 
Naturally the feature of $\delta(S)$ trajectories depends on the weighting pattern of Fourier modes, 
i.e. properties of the window function \citep[see examples in][]{BondEtal1991}. 

If the window function is sharp in k-space (a top-hat function spanning from $k=0$ to $k\sim 1/R$), 
the increment $\delta(S+dS)-\delta(S)$ of a step from $S$ to $S+dS$ comes from a new set 
of Fourier modes in a thin shell of
$(k, k+dk)$. Phases of $\delta_k$ are uniformly distributed in $[0, 2\pi]$, the sum 
$\sum_k^{k+dk}\delta_k$ is a random Gaussian variable and uncorrelated 
with previous increments \citep{BondEtal1991, LaceyCole1993}. This is exactly a Brownian random walk.
If we define $Q(\delta, S)$ as the number density of trajectories at $S$ within $(\delta, \delta+d\delta)$, 
the Brownian random walk satisfies a simple diffusion equation 
\begin{equation}
\frac{\partial Q}{\partial S}=\frac{1}{2}\frac{\partial^2Q}{ {\partial \delta}^2}\ ,
\label{BmDE}
\end{equation}
and $S=0$, $\delta(S)=0$. In absence of barrier, we have solution
\begin{equation}
Q(\delta,S)=\frac{1}{\sqrt{2\pi S}}\exp\left( -\frac{\delta^2}{2S} \right)\ .
\end{equation}

According to \citet{Chandrasekhar1943}, a
trajectory $\delta(S)$ reaches the barrier $\delta_c$ at $S$ has equal probability to 
walk above or below the barrier, therefore the solution of Eq.~\ref{BmDE} with an absorbing 
barrier boundary condition is
\begin{equation}
Q(S, \delta, \delta_c)=\frac{1}{\sqrt{2\pi S}}\left[e^{-\delta^2/2S}-
e^{-(\delta-2\delta_c)^2/2S}\right]\ .
\label{BmQ}
\end{equation}
The probability of a trajectory absorbed by the barrier $\delta_c$ must equal to the reduction of 
trajectories survived below the barrier in interval $(S, S+dS)$,
\begin{equation}
f_S(S, \delta_c)=-\frac{\partial}{\partial S}\int^{\delta_c}_{-\infty} Q d\delta\ .
\end{equation}
Substituting Eq.~\ref{BmDE} and \ref{BmQ} into the above equation gives
\begin{equation}
f_S(S,\delta_c)dS=\frac{\delta_c S^{-3/2}}{\sqrt{2\pi}}\exp\left( -\frac{\delta_c^2}{2S}\right) dS \ ,
\end{equation}
which is the fraction of mass associated with halos in the range of $S$ and consequently $M$. So the comoving
number density of halos of mass at epoch $z$ is simply
\begin{equation}
\frac{dn}{dM}dM=\frac{\bar{\rho}}{M}f_S\left| \frac{dS}{dM} \right| dM=
\frac{\bar{\rho}}{M^2}f_{PS}(\sigma)\left| \frac{d\ln \sigma}{d\ln M} \right| dM
\end{equation}
where 
\begin{equation}
f_{PS}(\sigma)=\sqrt{\frac{2}{\pi}} \frac{\delta_c}{\sigma}\exp \left( -\frac{\delta_c^2}{2\sigma^2} \right)\ .
\end{equation}
This is the well-known Press-Schechter mass function.

%===============================================================================

\section{the fractional excursion set theory} 
\subsection{motivation}
It is clear that the validity of the Brownian random motion prescription to the 
trajectory of $\delta(S)$ is guaranteed by the sharp k-space filtering. 
Lack of correlation between the new increment with any previous steps delimits the 
Markov nature of the Brownian motion. In context of structure formation, it means that 
the formation of halos at small scales is not correlated with the density fluctuation smoothed
at large scales, henceforth halo formation is completely independent of environment.

If we choose a different smoothing window function such as a Gaussian or a top-hat in real space, 
$\delta(S+dS)$ contains the same set of $\delta_k$ as $\delta(S)$ though in the summation
each Fourier mode is weighted differently by the window function. In this circumstance 
$\delta(S)$ is apparently correlated with earlier steps, which can not be described by the
Brownian random walk formalism any longer. In general there is no analytical solution to these types of
walks with correlation \citep{BondEtal1991}.

Recently with the emergence of high resolution simulations, it has been revealed that the 
formation history and properties of halo, especially of small mass, are modulated significantly
by halos' large scale environment \citep{ShethTormen2004,GaoEtal2005, HarkerEtal2006, WechslerEtal2006}. 
Therefore there must be considerable influence from the 
mass accretion at large scales on the amplitude of density fluctuation smoothed at small scales, i.e.
$\delta(S)$ is correlated with $\delta(S'<S)$ even the window function is sharp in k-space.

Either mathematically or physically, the walk of $\delta(S)$ of a realistic density field 
is some kind of random motion with correlated steps, which is obviously not a Brownian random motion, 
rather, is partly random and partly deterministic. Walks like this with ``fractional'' randomness, 
fortunately, are objects that the fractional Brownian motion (FBM) is designed to score.

%----------------------------------------------------------------------
\subsection{the fractional Brownian motion}
The FBM is a generalization of the normal Brownian random walk introduced by \citet{MandelbroltNess1968}. 
FBM, though not well-known in astronomy community, has been widely used to model geometry and growth 
of many types of rough surfaces in nature like mountain terrain, clouds, percolation and 
diffusion-limited aggregation. 
Interestingly it also finds its application in financial market \citep[c.f.][]{Meakin1998}.

Formally, with index $\alpha$ ($0< \alpha < 1$), a FBM is defined as a random process $X(t)$ on
some probability space such that: 
\begin{enumerate}
\item with probability 1, $X(t)$ is continuous and $X(0)=0$; 
\item for any $t\ge 0$ and $h> 0$, the increment $X(t+h)-X(t)$ follows a normal distribution with mean
zero and variance $h^{2\alpha}$, so that
\begin{equation}
P\left( X(t+h)-X(t)\le x \right)=\frac{h^{-\alpha}}{\sqrt{2\pi}}\int_{-\infty}^x 
e^{-u^2/2h^{2\alpha} }du\ .
\end{equation}
\end{enumerate}
If $\alpha=1/2$, FBM backs to the normal Brownian motion \citep[c.f.][]{Feder1988}.

The index $\alpha$ is named the Hurst exponent, which is used originally in the 
rescaled range analysis (R/S analysis) to portray scaling behaviors of time series. 
It has strong connection with the fractal dimensions of time series or spacial structures, but 
the exact relation is case dependent \citep{Meakin1998}. Here, $\alpha$ tells us
how strongly correlated the step increment is with previous steps. The trajectory of a FBM with 
smaller index $\alpha$ is more noisy than that of a FBM with higher index, so sometimes
$\alpha$ is called the roughness exponent.

It is very interesting that the FBM has infinitely long-run correlations. For instance,
the past increments $X(0)-X(-t)$ are correlated with future increments $X(t)-X(0)$: as
$X(0)=0$, the correlation function of the ``past'' and ``future'' is
\begin{equation}
C(t)=\frac{\langle -X(-t)X(t)\rangle}{\langle X^2(t)\rangle}=2^{2\alpha-1}-1
\end{equation}
which is invariant with the ``time'' $t$ and only vanishes when $\alpha=1/2$! This is an impressive
feature of FBM, which leads us to classify FBM into two types: 
\begin{enumerate}
\item {\em persistence} FBM with $\alpha>1/2$, which means that an increasing trend in past will result 
in an increasing trend in future for arbitrary large $t$, i.e. a positive feedback process; 
\item {\em anti-persistence} FBM with $\alpha<1/2$, which refers to an increasing trend in past will lead to a 
decreasing trend in  future, i.e. a negative feedback.
\end{enumerate}

It might help understanding characteristics of FBM to know the 
generation methods of FBM. To simulate a 1-dimensional FBM, the simplest method is 
\begin{equation}
X(t)=\frac{G(t)+\sum_{s=t-n}^{t-1}\left( t-s \right)^{\alpha-1/2}
G(s)}{\Gamma(\alpha+1/2)}\ ,
\end{equation}
in which $G(t)$ and $G(s)$ are uncorrelated random numbers extracted from a normal distribution with zero mean
and unity variance, $n$ is a practical cut-off number which shall be as large as possible.

To generate a $(d+1)$-dimensional surface of FBM by Fourier transformation, in the first instance we
place a grid in Fourier space, and fill the grid with complex numbers $\delta({\bf k})$ with 
Gaussian distributed amplitudes and random phases. Spatial correlation is introduced by
\begin{equation}
\delta'({\bf k})=k^{-(\alpha+d/2)}\delta({\bf k})\ .
\label{mkfbm}
\end{equation}
Then Fourier transformation of the random field $\delta'({\bf k})$ will give a self-affine surface modelled by
FBM.

%-------------------------------------------------------------------------
\subsection{the fractional excursion set theory}
The number density of trajectories $Q_\alpha(\delta, S)$ of fractional Brownian motion 
with index $\alpha$ obeys the diffusion equation \citep[c.f.][]{Lutz2001},
\begin{equation}
\frac{\partial Q_\alpha}{\partial S}=\alpha S^{2\alpha-1}\frac{\partial^2Q_\alpha}{{\partial \delta}^2}\ ,
\end{equation}
which has solution in absence of barrier
\begin{equation}
Q_\alpha(\delta,S)=\frac{S^{-\alpha}}{\sqrt{2\pi}}\exp\left( -\frac{\delta^2}{2S^{2\alpha}} \right)\ .
\end{equation}

Apparently the distribution of $\delta(S)$ at $S$ is still a Gaussian, the argument 
of \citet{Chandrasekhar1943} shall be valid, thus the solution under boundary condition of a fixed
absorbing barrier $\delta_c$ is
\begin{equation}
Q_\alpha(S, \delta, \delta_c)=\frac{S^{-\alpha}}{\sqrt{2\pi}}\left[e^{-\delta^2/2S^{2\alpha}}-
e^{-(\delta-2\delta_c)^2/2S^{2\alpha}}\right]\ .
\end{equation}

After a straightforward and tedious calculation, the halo mass function is
\begin{equation}
\frac{dn}{dM}dM=\frac{\bar{\rho}}{M^2}f_\alpha(\sigma)\left| \frac{d\ln \sigma}{d\ln M} \right| dM
\end{equation}
with the kernel 
\begin{equation}
f_\alpha(\sigma)=\frac{4\alpha}{\sqrt{2\pi}} \frac{\delta_c}{\sigma^{2\alpha}}
\exp \left( -\frac{\delta_c^2}{2\sigma^{4\alpha}} \right)\ .
\label{nmf}
\end{equation}
It is easy see that it is the PS function when $\alpha=1/2$. For comparison, we reproduce
the kernel of the Sheth-Tormen mass function here, 
\begin{equation}
f_{ST}(\sigma)=A\sqrt{\frac{2a}{\pi}} \left[ 1+\left( \frac{\sigma^2}{a\delta_c^2}\right)^p \right]
\frac{\delta_c}{\sigma}\exp\left( -\frac{a\delta_c^2}{2\sigma^2} \right)\ ,
\end{equation}
where $A=0.3222$, $a=0.707$ and $p=0.3$.

$f_\alpha$ of different Hurst exponent $\alpha$ in comparison with PS and ST formulas are displayed
in Fig.~1. The mass function from {\em persistence} FBM is very different with that of 
{\em anti-persistence} FBM. It appears that the ST function is in good 
agreement with our new mass function of index $\alpha=0.435$ in large mass regime of 
$\ln\sigma^{-1}>\sim0.3$ beyond which the mass function is very sensitive to the choice of $\alpha$. 
Considering the fact that ST mass function is obtained from fitting to simulations and has good accuracy
in large mass regime, an immediate conclusion is that trajectories $\delta(S)$ in our universe is 
actually {\em anti-persistent}.

In small mass regime, the dependence on $\alpha$ of $f_\alpha$ is relatively weak. If $\alpha<0.5$
the new mass function predicts less number of halos than the PS formula by $\sim10-20$\%, 
but up to $\sim30\%$ more than what ST function gives. It is very difficult and unreliable to 
resolve halos with mass lower than $10^8M_\odot$ in present days' simulations, we have to leave it 
to future to tell which mass function is better in small mass regime. A quick check indicates 
that $f_\alpha$ has very different shape with ST formula at $\ln\sigma^{-1}<\sim0.3$, we can 
only achieve good fit to $f_{ST}$ in range of $-0.5<\ln\sigma^{-1}<0.3$ with $\alpha\approx0.35$.

\begin{figure}
\resizebox{\hsize}{!}{\includegraphics{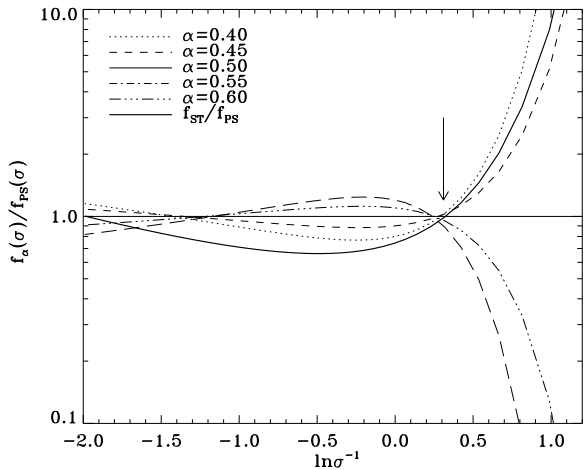}}
\caption{Ratios of the new mass function of different index $\alpha$ to the PS formula. The ratio
of the ST function to PS function is plotted in thick solid lines, which is in good agreement with
the new mass function of $\alpha\approx0.435$ in high mass regime.
The arrow indicates the region where all curves cross the line $f/f_{\rm PS}=1$.}
\end{figure}

%===============================================================================
\section{discussion}
The fractional Brownian motion of index $\alpha$ is introduced to construct the 
fractional excursion set theory. The new mass function computed with the theory is analytical and 
simple, of which the PS mass function is only a special case of $\alpha=1/2$. 
Comparison with the ST function nurtured by N-body simulations demonstrates
the excellent performance of the new mass function.

In Fig.~1 it is observed that high mass halo abundance is very sensitive to the value of 
$\alpha$, the high mass halo abundance observed can be potentially a very powerful tool to 
detect the non-Gaussianity of the initial density fluctuation field: non-Gaussianity will change
the correlation between walking steps of $\delta(S)$ and therefore modify the $\alpha$ effectively.

The success of applying FBM formalism to model structure formation is attributed to 
the inclusion of the correlation between density fluctuations at different scales. 
The correlation strength characterized by the Hurst exponent $\alpha$ could be resulted from
properties of window function and the intrinsic correlation of the cosmic density
field. We know that a non-sharp filtering in k-space will induce correlation \citep{BondEtal1991}, but 
are unclear how $\alpha$ changes with features of the window function. 
More of interests is the relation between $\alpha$ and the power spectrum of density field.
The generation method Eq.~\ref{mkfbm} provides some clues, however there is the complication that
the scaling of trajectory $\delta(S)$ is founded relative to the variance $\sigma^2$, 
not the physical scale $R$. Numerical experiments with scale free simulations shall be able
to improve our understanding effects of window function and power spectrum on $\alpha$. 

In this work only the mass function is computed. In principle, the fractional
excursion set theory may have many applications, for example, those works 
of \citet{LaceyCole1993}, \citet{MoWhite1996} and \citet{ZhangHui2006}
can all be revisited with FBM. Since $\alpha$ denotes the correlation of $\delta$ at different scales,
the subsequently calculated halo merger rate and merger history is marked with
the stamp of large scale environment on halo formation at small scales. We might
be able to explain the halo clustering dependence on halo formation history and environment
\citep{GaoEtal2005, WechslerEtal2006}.

The new halo mass function is obtained assuming spherical collapse. To improve
the accuracy of the model, ellipsoidal collapse has to be taken into account. The poor performance
of Eq.~\ref{nmf} of $\alpha\approx0.435$ in low mass regime (see Fig.~1) is very likely due to our
simplification of adopting the spherical collapse model. Essentially to calibrate the ellipsoidal
collapse, we replace the fixed barrier $\delta_c$ with a moving barrier $B(S)$ as in 
\citet{ShethTormen2002} and \citet{DelPopolo2006a, DelPopolo2006b}, and then solve 
the diffusion equation with the new boundary condition. Technique details and comparison with
simulations will be presented elsewhere (Fosalba \& Pan, in preparation).

%===============================================================================
\section*{Acknowledgement}
The author thanks the kind help in FBM from Shaohua Hu, enjoys the fruitful discussion with 
Yanchuan Cai, Xuelei Chen, Longlong Feng, Weipeng Lin, Luis Teodoro, Istvan Szapudi, and 
appreciates valuable suggestions from Pablo Fosalba, Yipeng Jing, Pengjie Zhang and the referee. JP 
is supported by a preliminary funding from PMO under scheme of the One Hundred Talent 
program of China Academy of Science. 
%\bibliography{cc}
%\bibliographystyle{mn2e}

\input{fres.bbl}
\end{document}

%% file: aas_journals.tex
\def\aj{{AJ}}                   % Astronomical Journal
\def\apj{{ApJ}}                 % Astrophysical Journal
\def\aap{{A\&A}}                % Astronomy and Astrophysics
\def\mnras{{MNRAS}}             % Monthly Notices of the RAS